\documentclass[preprint,tightenlines,aps,floats,nofootinbib]{revtex4}
\usepackage{epsf}
\usepackage{graphicx}
\begin{document}          

\vskip 2in
\hskip 5.2in \hbox{\bf YITP-SB-10-30}

\title{\vspace*{0.7in}
Four Generations, Higgs Physics and the MSSM}

\author{S.~Dawson$^{a}$ and P.~Jaiswal$^{a,b}$}

\affiliation{
$^a$Department of Physics, Brookhaven National Laboratory, 
Upton, NY 11973, USA \\
$^{b}$Yang Institute for Theoretical Physics, Stony Brook University, Stony Brook, NY 11790, USA
\vspace*{.5in}}

\begin{abstract}
We consider the effects of a fourth generation of chiral fermions
within the MSSM.  Such a model offers the possibility of having the
lightest neutral Higgs boson significantly heavier than in the three
generation MSSM.  The model is highly constrained by precision electroweak
data, along with Higgs searches at the Tevatron.  In addition, 
the requirements of perturbative unitarity
and  direct searches for heavy quarks imply that
the four generation MSSM is only consistent for $\tan\beta\sim 1$ and 
highly tuned $4^{th}$ generation  fermion masses. 
\end{abstract}

\maketitle
\thispagestyle{empty}
\newpage
\pagestyle{plain}

\section{Introduction}
The Standard Model offers no clue as to why only three generations of
chiral fermions are observed.   It is thus natural to consider the consequences
of a fourth family of heavy fermions\cite{Holdom:2007nw,Soni:2010xh}.  
The allowed parameter space for a fourth generation 
is severely restricted
by experimental searches, by precision electroweak measurements,
and by theoretical constraints from the requirements implied by
the perturbative
 unitarity of heavy fermion scattering amplitudes
and the perturbativity of the Yukawa coupling constants at high energy.

A model with a fourth generation contains charge $2/3$ and $-1/3$ quarks,
$t^\prime$ and $b^\prime$, and a charged lepton, $e^\prime$, with its
associated neutrino, $\nu^\prime$.
Tevatron searches for direct production of a $b^\prime$ \cite{Aaltonen:2009nr}
imply $m_{b^\prime}>338 ~GeV$, assuming $b^\prime\rightarrow Wt$, and
$m_{t^\prime} >335~GeV$, assuming 
$t^\prime\rightarrow W q$, with $q=d,s,b$\cite{Lister:2008is}.
Relaxing the mixing assumptions changes the limits somewhat, but the $b^\prime$
limits vary by less than 20$\%$, while the $t^\prime $ limits increase
in some mixing scenarios to 
$m_{t^\prime}> {\cal {O}}(400~GeV)$\cite{Flacco:2010rg}.  
In all cases, a fourth generation quark is excluded up to a mass of
${\cal O}(300~GeV)$.
We consider 4th generation neutrinos heavier than $M_Z/2$, so there
is no constraint from the invisible $Z$ width.  From direct production
searches for
 $e^\prime$ and $\nu^\prime$ at LEPII, 
there is a limit of ${\cal {O}}(100~GeV)$ on
the masses of  4th generation
charged leptons and unstable neutrinos.  Current bounds on 4th generation
Standard Model like fermions are reviewed in 
Ref. \cite{Kribs:2007nz,Hung:2007ak,Eberhardt:2010bm,Bobrowski:2009ng}.
We will typically consider $4^{th}$
generation lepton masses greater than $\sim 200~GeV$ and quark
masses greater than $\sim 300-400~GeV$, which are safely above
direct detection bounds.  Furthermore, we will neglect CKM mixing
between the $4^{th}$ generation and the lighter $3$ 
generations\cite{Chanowitz:2010bm}.

Precision electroweak measurements place strong constraints on the the allowed
fermion masses of a $4^{th}$ generation, 
but it is possible to arrange the masses such 
that cancellations
occur between the contributions of the heavy leptons and quarks.
By carefully tuning the fourth
generation fermion masses, the Higgs boson can be as  heavy as 
$M_h\sim 600~GeV$\cite{Haller:2010zb,gfit,Flacher:2008zq,Kribs:2007nz}.
In a four generation model, therefore, Higgs 
physics can be significantly altered
from that of the Standard Model:  Higgs production from gluon fusion
is enhanced by a factor of 
roughly $9$\cite{Anastasiou:2010bt}, and the Higgs branching ratio
to $2$ gluons is similarly enhanced\cite{Kribs:2007nz}.
  The D0 experiment has recently
excluded a SM-like Higgs mass between $131~GeV$
 and $204~GeV$ produced from gluon fusion in
a four generation scenario\cite{Aaltonen:2010sv}.

It is interesting to consider scenarios with heavy fermions and a neutral
Higgs boson heavier than  expected from Standard Model electroweak
fits.  A model of this type is
the MSSM with a fourth generation of chiral fermions (4GMSSM).  This
model has a number of interesting features.  Since the mass-squared of  the lightest
Higgs boson in the MSSM receives corrections proportional to the (mass)$^4$
of the heavy fermions, it is potentially possible to significantly
increase the lightest
Higgs boson mass in the four generation version of the 
MSSM\cite{Litsey:2009rp}. 
In general, a $4^{th}$ generation of heavy quarks can contribute to
electroweak baryogenesis\cite{Hou:2008xd,Kikukawa:2009mu}
 and Ref. \cite{Fok:2008yg}
 argues
that the 4GMSSM with $\tan\beta\sim 1$ can yield a first order
electroweak phase transition for $4^{th}$ generation quark
and squark masses just beyond the current Tevatron search bounds. 

We discuss the features of the model in Section \ref{modelsec},
and derive unitarity
constraints on the fermion masses in Section \ref{unitsec}. 
In the 4GMSSM, 
these constraints
can be quite different from those of 
the four generation version of the Standard Model\cite{Chanowitz:1978mv}.
Section \ref{stusec} contains limits on the four generation
MSSM from precision electroweak measurements. 
Section \ref{conc} contains some conclusions.

\section{The Model}
\label{modelsec}
We consider an $N=1$ supersymmetric model 
which is an exact replica of the $3$ generation MSSM except that it
contains
a 4th generation of chiral superfields described by
the superpotential\cite{Godbole:2009sy,Murdock:2008rx,Hashimoto:2010at,Gunion:1995tp,Carena:1995ep}
\begin{equation}
W_4=
 \lambda_{t^\prime}{\hat \psi_4} ({\hat t^\prime})^c {\hat H_2}
+\lambda_{b^\prime}{\hat \psi_4} ({\hat b^\prime})^c {\hat H_1}
+\lambda_{e^\prime} {\hat l_4} ({\hat e^\prime})^c {\hat H_1}
+\lambda_{\nu^\prime} {\hat l_4} ({\hat \nu^\prime})^c {\hat H_2} 
\, ,
\label{super4}
\end{equation}
where ${\hat \psi_4}$ is the $4^{th}$ generation
$SU(2)_L$ quark and squark doublet superfield,
${\hat l_4}$ is the $4^{th}$ generation 
$SU(2)_L$ lepton and slepton doublet superfield,
and ${\hat H}_i$ are the $SU(2)_L$ Higgs superfields.  Similarly,
${\hat t^\prime},{\hat b^\prime}, {\hat e^\prime}$
and ${\hat \nu^\prime}$
are the $4^{th}$ generation
superfields corresponding to the right-handed fermions.
We assume no mixing between $W_4$ and the superpotential of the
$3$ generation MSSM\footnote{Limits on the 4 generation
 Standard Model
suggest that the mixing between the $3^{rd}$ and $4^{th}$
generation is restricted to be small, 
$\theta_{34} < .1$\cite{Kribs:2007nz,Chanowitz:2010bm} .}. 
The new particles in the 4GMSSM
are  the $4^{th}$ generation quarks and leptons (including
a right-handed heavy neutrino), along with their associated scalar partners.
We assume that the $4^{th}$ generation neutrino receives a Dirac mass,
although our conclusions are relatively insensitive to these
assumptions.

The Higgs sector is identical to the $3$ generation MSSM and
consists of $2$ neutral scalars, $h$ and $H$, a pseudo-scalar, $A$,
and a charged scalar, $H^\pm$. The Higgs
Yukawa couplings of $t^\prime$,
$b^\prime$,$e^\prime$ and $\nu^\prime$ are,
\begin{eqnarray}
\lambda_{t^\prime}& =&{m_{t^\prime}\sqrt{2}\over v\sin\beta}
\quad\quad
\lambda_{b^\prime} ={m_{b^\prime}\sqrt{2}\over v\cos\beta}
\nonumber \\
\lambda_{e^\prime}& =&{m_{e^\prime}\sqrt{2}\over v\cos\beta}
\quad \quad
\lambda_{\nu^\prime} ={m_{\nu^\prime}\sqrt{2}\over v\sin\beta}
\, ,
\end{eqnarray}
where $tan\beta$ is the usual ratio of Higgs vacuum
expection values\cite{Gunion:1989we}.
Because of the large masses of the $4^{th}$ generation
fermions which are required
in order to satisfy restrictions from
the experimental searches, 
the Yukawa couplings quickly  become non-perturbative.
Requiring perturbativity at the weak scale, a strong bound
comes from the restriction
$\lambda_{b^\prime}^2<4\pi$ which implies\cite{DePree:2009ed},
\begin{equation}
\tan\beta < 
\sqrt{2 \pi \biggl({v\over m_{b^\prime}}\biggr)^2-1}\sim 1.8
\, ,
\end{equation}
for $m_{b'}\sim 300~GeV$.
The evolution of the Yukawa couplings above the
weak scale has been studied in 
Refs. \cite{Murdock:2008rx,Godbole:2009sy,Hashimoto:2010at}
with the conclusion that it is not possible for the 4GMSSM to
be perturbative above scales on the order of $10-1000~TeV$.  The
4GMSSM thus leads to a picture with an intermediate scale of physics
such as that present in  gauge mediated SUSY models. 

\begin{figure}[t]
\begin{center}
\includegraphics[scale=0.6]{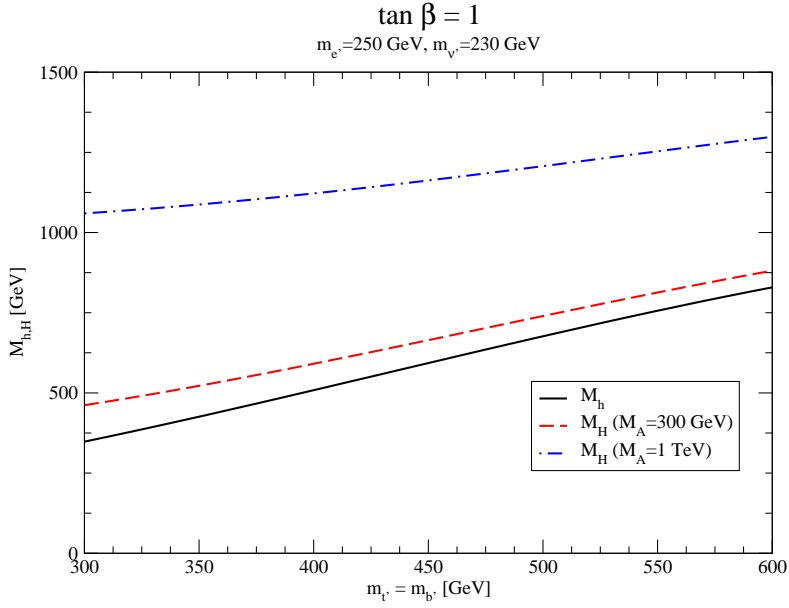}
\caption[]{Predictions for the neutral Higgs boson 
masses in the four generation
MSSM.  The squarks and sleptons 
are assumed to 
have degenerate masses of
$1~TeV$.
The mass of the lighter
Higgs boson, $M_h$, is insensitive to the value of $M_A$.
(Not all masses shown here are allowed by the restrictions of
perturbative unitarity and electroweak precision measurements, as
discussed in Sects. \ref{unitsec} and \ref{stusec}.)
}
\label{fg:mh}
\end{center}
\end{figure}

In the 4GMSSM, the lightest Higgs boson mass has an upper bound which 
receives large corrections proportional to the 
$4^{th}$ generation  fermion masses.
The masses of the neutral Higgs bosons can therefore be significantly
heavier than in the case of
 the $3$ generation MSSM and  are shown in Fig. \ref{fg:mh}
for $\tan\beta=1$ and representative $4^{th}$ generation
masses\cite{Litsey:2009rp}.
The dominant contributions to the neutral Higgs masses in the
4GMSSM are given
in Appendix A\cite{Haber:1993an,Haber:1990aw,Ellis:1990nz,Heinemeyer:1998kz}.  

\section{Tree Level Unitarity}
\label{unitsec}

\begin{figure}[t]
\begin{center}
\includegraphics[scale=0.6]{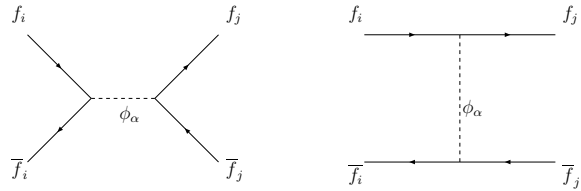}
\caption[]{Feynman diagrams contributing to $f_i
{\overline f}_i\rightarrow  f_j
{\overline f}_j$ in the high energy limit. $\phi_\alpha$ is
a scalar, pseudo-scalar, or Goldstone boson.
}
\label{fg:feyn}
\end{center}
\end{figure}

Chiral fermions have an upper bound on their masses from the requirement
of perturbative unitarity of fermion anti-fermion scattering at high 
energy,
originally derived in Ref. \cite{Chanowitz:1978mv}.
In the MSSM, the unitarity bounds on heavy fermions can be quite
different from those of the Standard Model, 
due to the effects of the additional
scalars present in the MSSM, and also
to  the different fermion Yukawa couplings in the
MSSM relative to those of the Standard Model.

Consider an $SU(2)_L$ doublet of heavy left-handed fermions, along
with their corresponding right-handed fermion partners,
\begin{equation}
\psi_L=\left(\begin{array}{c} f_1\\
f_2\end{array}\right)_L\, , \,\, f_{1R}, f_{2R}\, ,
\label{fermdef}
\end{equation}
with masses $m_1$ and $m_2$.
 At high energy, $\sqrt{s}>>m_i$, the scattering amplitudes can be most 
conveniently written in terms of helicity amplitudes.  The positive and
negative helicity spinors are 
$u_\pm(p)=P_{L,R}u(p)$ and $v_\pm
=P_{L,R}v(p)$, where $P_{L,R}={1\over 2}(1\mp \gamma_5)$.  The fermions
interact with the scalars of the MSSM and the Goldstone bosons of
electroweak symmetry breaking via the interactions,
\begin{equation}
L={\overline f}_i
\biggl(a_L^{i\alpha}P_L+a_R^{i\alpha}P_R\biggr)f_i
\phi_\alpha^0 +\biggl\{
{\overline f}_1
\biggl(a_L^{12\alpha}P_L+a_R^{12\alpha}P_R\biggr)f_2
\phi_\alpha^+ + h.c.\biggr\}\, ,
\end{equation}
where $\phi_\alpha^0$ and $\phi_\alpha^\pm$ are generic neutral and
charged scalars.  
 
The scattering of ${\overline f_i}^\lambda 
f_i^{\overline{\lambda}}\rightarrow
{\overline f_j}^{\lambda^\prime} f_j^{\overline{\lambda^\prime}}$ 
can be found using  the Goldstone Boson equivalence
theorem to obtain the high energy limits
(where $\lambda$ are the helicity indices). The Feynman
diagrams are shown in Fig. \ref{fg:feyn}.  
In the $s-$ channel,
the contribution from neutral
scalar or pseudo-scalar exchange, $\phi_\alpha^0$,
to the generic amplitude for $
f_i {\overline f}_i \rightarrow f_j
{\overline f_j}$  in the high energy limit is, 
\begin{equation}
M_s=\overline{u_{\lambda^\prime}}(p_3)(a_L^{j\alpha} P_L
+a_R^{j\alpha} P_R)
v_{\overline {\lambda^\prime}}(p_4)
~\overline{v_{{\overline{\lambda}}^\prime}}(p_2)(a_L^{i\alpha} P_L
+a_R^{i\alpha} P_R)u_{\lambda}(p_1)\, .
\end{equation} 
The high energy limits of the 
helicity amplitudes from the $s-$channel contributions are thus,
\begin{eqnarray}
M_s(++\rightarrow ++)&=& +a_L^{i\alpha} a_R^{j\alpha} s \nonumber \\
M_s(++\rightarrow --) &=& -a_L^{i\alpha} a_L^{j\alpha} s \nonumber \\
M_s(--\rightarrow ++) &=& -a_R^{i\alpha} a_R^{j\alpha} s \nonumber \\
M_s(--\rightarrow --)&=& +a_R^{i\alpha} a_L^{j\alpha} s \, ,
\label{samp}
\end{eqnarray} 
where $s=(p_1+p_2)^2$, $t=(p_1-p_3)^2$,
and  we have assumed $s>>m_i^2,M_\phi^2,M_W^2$, and $ M_Z^2$.

Similarly, the high energy limit of the
amplitude resulting from the exchange of a  scalar or
pseudo-scalar in the $t-$ channel is,  
\begin{equation}
M_t=
\overline{u_{\lambda^\prime}}(p_3)
(a_L^{ij\alpha} P_L
+a_R^{ij\alpha} P_R)
u_{\lambda}(p_1)~
\overline{v_{\overline{\lambda}}}(p_2)(a_L^{ij\alpha} P_L
+a_R^{ij\alpha} P_R)
v_{{\overline{\lambda^\prime}}}(p_4)\, ,
\end{equation}
which yields  the helicity  amplitudes,
\begin{eqnarray}
M_t(++\rightarrow --)&=& -(a_L^{ij\alpha})^2 t \nonumber \\
M_t(--\rightarrow ++) &=& -(a_R^{ij\alpha})^2 t \nonumber \\
M_t(+-\rightarrow -+) &=& + a_R^{ij\alpha} a_L^{ij\alpha} t \nonumber \\
M_t(-+\rightarrow +-)&=& + a_L^{ij\alpha} a_R^{ij\alpha} t \, .
\label{tamp}
\end{eqnarray} 
(We have assumed that all couplings are real).

From the results in Eqs. \ref{samp}
and \ref{tamp}, it is straightforward to read off the
contributions to the partial wave amplitudes for 
a specific model.  The MSSM couplings of the
fermions to the scalars can be found in
 Ref. \cite{Gunion:1989we}, for example.
First consider the scattering of ${\overline f}_1 f_1\rightarrow
{\overline f_1} f_1$ in the 4GMSSM.  
In the $s-$ channel, $h,H,A$, and $G^0$ contribute
 and 
their contributions are
 listed in Table \ref{tb:sch1}, while the 
$t-$ channel contributions are shown in Table
\ref{tabtch1}\footnote{We have defined 
$s_\beta=\sin\beta$, $c_\beta=\cos\beta$,
$s_\alpha=\sin\alpha$ and $c_\alpha=\cos\alpha$.  The mixing
in the neutral Higgs sector is described by the angle $\alpha$ which
is defined in Appendix A and in Ref. \cite{Gunion:1989we}.}.
 It is apparent that there are many cancellations 
between the various contributions that are not present in the
Standard Model.  The amplitudes for  ${\overline f_2} f_2\rightarrow
{\overline f_2} f_2$ are found by making the replacments $m_1\rightarrow m_2$, $\beta
\rightarrow \beta+{\pi\over 2}$, $\alpha\rightarrow \alpha -{\pi\over 2}$.

\begin{table}[t]
\begin{tabular}{|c|c|c|c|c|}
\hline
$\lambda {\overline \lambda}
\rightarrow \lambda^\prime {\overline \lambda}^\prime$
& $M_h$ & $M_H$ & $M_A$
&   $M_{G^0}$\\
\hline
$++\rightarrow ++$ &
$ -{c_\alpha^2\over s_\beta^2}$ &
$-{s_\alpha^2\over s_\beta^2}$ &
$-\cot^2\beta$ & $-1$\\
$++\rightarrow --$ & 
$+{c_\alpha^2\over s_\beta^2}$ &
$+{s_\alpha^2\over s_\beta^2}$ &
$-\cot^2\beta$ & $-1$\\
$--\rightarrow ++$ & 
$+{c_\alpha^2\over s_\beta^2}$ &
$+{s_\alpha^2\over s_\beta^2}$ &
$-\cot^2\beta$ & $-1$\\
$--\rightarrow --$ & 
$-{c_\alpha^2\over s_\beta^2}$ &
$-{s_\alpha^2\over s_\beta^2}$ &
$-\cot^2\beta$ & $-1$\\
 \hline
\end{tabular}
\caption{Contributions from $s-$ channel exchange
of $h,H,A$, and $G^0$ to helicity scattering amplitudes
for ${\overline f_1} f_1\rightarrow
{\overline f_1} f_1$ in the high energy limit
of the 4GMSSM. The contributions
given in the table must be  multiplied by $\sqrt{2} G_F m_1^2$.}
\label{tb:sch1}
\end{table}
\noindent

\begin{table}[t]
\begin{tabular}{|c|c|c|c|c|}
\hline
$\lambda {\overline \lambda}
\rightarrow \lambda^\prime {\overline \lambda}^\prime$
& $M_h$ & $M_H$ & $M_A$
&   $M_{G^0}$\\
\hline
$++\rightarrow --$ &
$+ {c_\alpha^2\over s_\beta^2}$ &
$+{s_\alpha^2\over s_\beta^2}$ &
$-\cot^2\beta$ & $-1$\\
$--\rightarrow ++$ & 
$+{c_\alpha^2\over s_\beta^2}$ &
$+{s_\alpha^2\over s_\beta^2}$ &
$-\cot^2\beta$ & $-1$\\
$+-\rightarrow -+$ & 
$-{c_\alpha^2\over s_\beta^2}$ &
$-{s_\alpha^2\over s_\beta^2}$ &
$-\cot^2\beta$ & $-1$\\
$-+\rightarrow +-$ & 
$-{c_\alpha^2\over s_\beta^2}$ &
$-{s_\alpha^2\over s_\beta^2}$ &
$-\cot^2\beta$ & $-1$\\
 \hline
\end{tabular}
\caption{Contributions from $t-$ channel exchange
of $h,H,A$, and $G^0$ to helicity scattering amplitudes
for ${\overline f_1} f_1\rightarrow
{\overline f_1} f_1$ in the high energy limit of the 4GMSSM.
The contributions
given in the table must be multiplied by $\sqrt{2} G_F m_1^2$.}
\label{tabtch1}
\end{table}
\noindent

Flavor changing fermion anti-fermion 
 scattering,  ${\overline f_1} f_1\rightarrow
{\overline f_2} f_2$, also yields interesting limits on
heavy fermion masses in the 4GMSSM.  The $s-$channel contributions
to the high energy limits of the helicity scattering amplitudes
are shown in Table \ref{tabs12}, and the $t-$ channel contributions from 
$H^+$ and $G^+$ exchange in Table \ref{tabt12}.

\begin{table}[t]
\begin{tabular}{|c|c|c|c|c|}
\hline
$\lambda {\overline \lambda}
\rightarrow \lambda^\prime {\overline \lambda}^\prime$
& $M_{h}$ & $M_H$ & $M_A$ & 
   $M_{G^0}$\\
\hline
$++\rightarrow ++$ &
$+ {\sin 2\alpha\over \sin 2 \beta}$&
$-{\sin 2\alpha\over \sin 2 \beta}$&
$-1$&$+1$\\
$++\rightarrow --$ &
$- {\sin 2\alpha\over \sin 2 \beta}$&
$+{\sin 2\alpha\over \sin 2 \beta}$&
$-1$&$+1$\\
$--\rightarrow ++$ &
$- {\sin 2\alpha\over \sin 2 \beta}$&
$+{\sin 2\alpha\over \sin 2 \beta}$&
$-1$&$+1$\\
$--\rightarrow --$ &
$+ {\sin 2\alpha\over \sin 2 \beta}$&
$-{\sin 2\alpha\over \sin 2 \beta}$&
$-1$&$+1$\\
 \hline
\end{tabular}
\label{tabs12}
\caption{Contributions from $s-$ channel exchange
of $h,H,A$ and $G^0$ to helicity scattering amplitudes
for ${\overline f_1} f_1\rightarrow
{\overline f_2} f_2$ in the high energy limit. An overall factor of
$\sqrt{2} G_F m_1 m_2$ is omitted.}
\end{table}

\begin{table}[t]
\begin{tabular}{|c|c|c|}
\hline
$\lambda {\overline \lambda}
\rightarrow \lambda^\prime {\overline \lambda}^\prime$
& $M_{H^+}$ & 
   $M_{G^+}$\\
\hline
$++\rightarrow --$ &
$+ m_1m_2$ &
$-m_1 m_2$ \\
$--\rightarrow ++$ & 
$+m_1 m_2$ &
$-m_1 m_2$ \\
$+-\rightarrow -+$ & 
$-m_2^2$ &
$-m_2^2$ \\
$-+\rightarrow +-$ & 
$-m_1^2$ &
$-m_1^2$ \\
 \hline
\end{tabular}
\caption{Contributions from $t-$ channel exchange
of $H^+$ and $G^+$ to helicity scattering amplitudes
for ${\overline f_1} f_1\rightarrow
{\overline f_2} f_2$ in the high energy limit. An overall factor of
$2\sqrt{2} G_F$ is omitted.}
\label{tabt12}
\end{table}
\noindent

\begin{figure}[t]
\begin{center}
\includegraphics[scale=0.6]{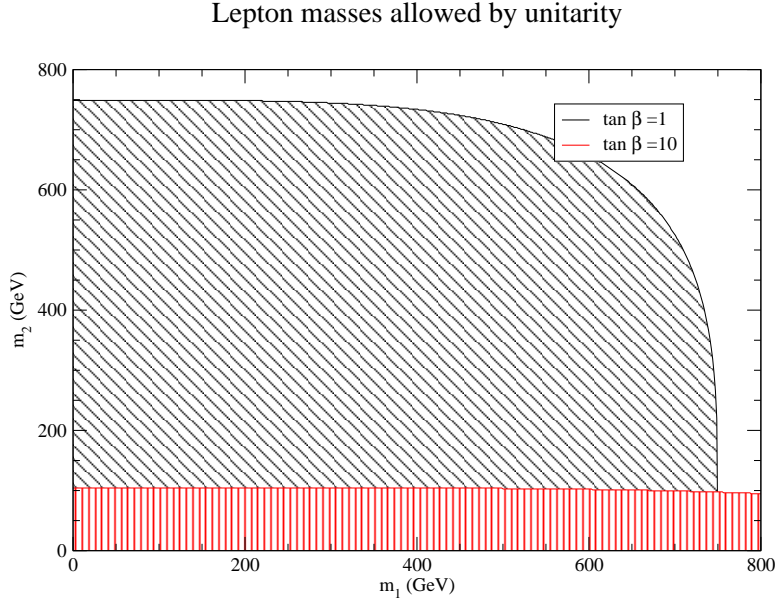}
\caption[]{Unitarity restriction on on a $4^{th}$ generation lepton
doublet in the 4GMSSM. The allowed 
region with the vertical (diagonal) cross- hatches
corresponds to $\tan\beta=10 (1)$. }
\label{fg:lepunit}
\end{center}
\end{figure}

Bounds on the fermion masses come from the coupled
channel $J=0$ partial wave amplitudes for
$f_i^{\lambda}
{\overline f_i}^{\overline{\lambda}}
\rightarrow f_j^{\lambda^\prime} 
{\overline f}_j^{\overline{\lambda^\prime}}$ 
\cite{Chanowitz:1978mv,Chanowitz:1978uj},
\begin{equation}
a_0={1\over 16\pi s}\int_{-s}^0 \mid M\mid\, ,
\end{equation}
where $\mid M\mid$ is the sum of the $s-$ and $t-$
channel helicity amplitudes given in the tables.
Perturbative unitarity requires
that the eigenvectors of the scattering matrix satisfy
 $\mid a_0\mid < 1$\cite{Lee:1977eg}. 
In the scattering basis, $f_1^+ {\overline f}_1^+, 
f_2^+ {\overline f}_2^+, f_1^- {\overline f}_1^-, 
f_2^- {\overline f}_2^-$, the high energy limit of 
the $J=0$ coupled partial
wave scattering matrix is,
\begin{equation}
\mid a_0
\mid \equiv B ={G_F\over 4\sqrt{2}\pi}
\left(
\begin{array}{cccc}
{m_1^2\over s_\beta^2} & 0 & 0 & 0\\
0 & {m_2^2\over c_\beta^2} & 0 & 0\\
0&0&{m_1^2\over s_\beta^2}&0\\
0&0&0&{m_2^2\over c_\beta^2}
\end{array}
\right)\, .
\label{unimat}
\end{equation}
Enforcing the unitarity condition, $\mid a_0\mid < 1$,
on the eigenvalues of Eq. \ref{unimat} gives 
the restrictions,
\begin{eqnarray}
m_1^2 & < & s_\beta^2{4\sqrt{2}\pi\over G_F}\nonumber \\
m_2^2 & < & c_\beta^2{4\sqrt{2}\pi\over G_F}\, .
\label{lepres}
\end{eqnarray}
A further interesting limit is found from
the coupled channel scattering of
the helicity amplitudes
$f_1^+ {\overline f}_1^-, 
f_1^- {\overline f}_1^+, f_2^+ {\overline f}_2^-, 
f_2^- {\overline f}_2^+$, with 
\begin{equation}
\mid a_0
\mid ={G_F\over 4\sqrt{2}\pi}
\left(
\begin{array}{cccc}
0& {m_1^2\over s_\beta^2} & 0      & {m_1^2\over s_\beta^2}  \\
{m_1^2\over s_\beta^2} & 0& {m_2^2\over c_\beta^2} & 0 \\
0& {m_2^2\over c_\beta^2}&0& {m_2^2\over c_\beta^2}\\
{m_1^2\over s_\beta^2}&0&{m_2^2\over c_\beta^2}&0
\end{array}
\right)\, .
\label{tch}
\end{equation}
Requiring the largest eigenvalue of Eq.~\ref{tch} to be $<~1$,
\begin{equation}
\lambda_{max}={G_F\over 4 \pi}\sqrt{
{m_1^4\over s_\beta^4} + {m_2^4\over c_\beta^4}}
<1\, .
\label{tch2}
\end{equation}
The bounds of Eqs \ref{lepres} and~\ref{tch2} are relevant for a heavy lepton
doublet
in the 4GMSSM and the allowed regions are shown in Fig. \ref{fg:lepunit}.
These bounds can be compared with the Standard Model bounds,
$m_{lepton}^2 < {4\sqrt{2}\pi\over G_F}=(1.2~TeV)^2$.  For $\tan\beta=1$,
the bound is reduced from the Standard
Model value to $m_{lepton}<750~GeV$. For $\tan\beta=10$, the
value of $m_2$ ($m_{\nu '}$) allowed
by unitarity is $m_{\nu '} < 100~GeV$, which is excluded by
experimental searches.

\begin{figure}[t]
\begin{center}
\includegraphics[scale=0.6]{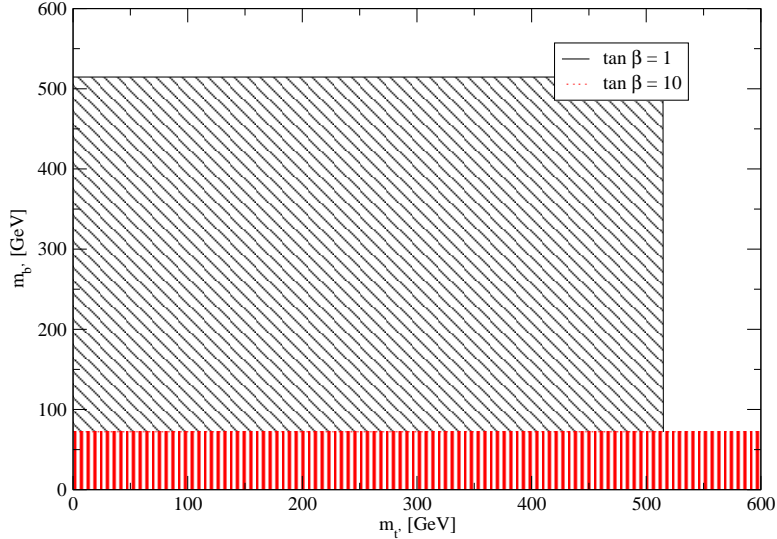}
\caption[]{Unitarity restriction  on a $4^{th}$ generation quark
doublet in the 4GMSSM. The allowed 
region with the vertical (diagonal) cross- hatches
corresponds to $\tan\beta=10 (1)$.}
\label{fg:quarkunit}
\end{center}
\end{figure}

The bounds on a heavy quark doublet in the 4GMSSM can be found by
considering the color neutral scattering amplitudes\footnote{
The logic is identical to Ref. \cite{Chanowitz:1978mv}.}. In the basis,
 $f_1^+ {\overline f}_1^+, 
f_2^+ {\overline f}_2^+, f_1^- {\overline f}_1^-, 
f_2^- {\overline f}_2^-$, the coupled channel scattering 
matrix for the $J=0$ partial wave
amplitude is a 12 $\times 12$ matrix of the form,
\begin{equation}
\mid a_0\mid \sim \left(
\begin{array}{ccc}
B & B & B\\
B & B & B\\
B & B & B
\end{array}
\right)\, ,
\label{uniquark}
\end{equation}
where the $3\times 3$ color neutral
 matrix $B$ is given in Eq. \ref{unimat}. Restricting
the eigenvectors to be less than $1$ gives the restrictions on
$4^{th}$ generation quark masses shown in Fig. \ref{fg:quarkunit},
\begin{eqnarray}
m_1^2 & < & s_\beta^2{4\sqrt{2}\pi\over 3 G_F}\nonumber \\
m_2^2 & < & c_\beta^2{4\sqrt{2}\pi\over 3 G_F}
\, .
\end{eqnarray}
It is apparent that the experimental bounds of $m_{t^\prime}>335~GeV$
and $m_{b^\prime}>338~GeV$ are close to violating perturbative
unitarity in the 4GMSSM with $\tan\beta=1$. For larger $\tan\beta$, 
the parameters are even more restricted. 

\section{Limits from Precision Electroweak Measurements}
\label{stusec}
The limits on the 4GMSSM  from precision electroweak measurements
can be studied
assuming that the dominant contributions
 are to the gauge
boson 2-point functions\cite{Peskin:1991sw,Altarelli:1990zd},
$\Pi_{XY}^{\mu\nu}(p^2)=\Pi_{XY}(p^2)g^{\mu\nu}+B_{XY}(p^2)p^\mu p^\nu$,
with $XY=\gamma\gamma, \gamma Z, ZZ$ and $W^+W^-$,
\begin{eqnarray}
\alpha S&=&
\biggl({4 s_W^2 c_W^2\over M_Z^2}\biggr)
\biggl\{ \Pi_{ZZ}(M_Z^2)- \Pi_{ZZ}(0)-
\Pi_{\gamma\gamma}(M_Z^2)
\nonumber \\ &&
-{c_W^2-s_W^2\over c_W s_W}\biggl(
\Pi_{\gamma Z}(M_Z^2)
-\Pi_{\gamma Z}(0)\biggr)\biggr\}\nonumber \\
\alpha  T &=&
 \biggl({ \Pi_{WW}(0)\over M_W^2}-{
\Pi_{ZZ}(0)\over M_Z^2}-{2 s_W
\over c_W }{\Pi_{\gamma Z}(0) \over M_Z^2}
\biggr)
\nonumber \\
\alpha U&=& 4 s_W^2\biggl\{
{ \Pi_{WW}(M_W^2)-\Pi_{WW}(0)\over M_W^2} -c_W^2
\biggl({ \Pi_{ZZ}(M_Z^2)-\Pi_{ZZ}(0)\over M_Z^2}\biggr)
\nonumber \\
&&-2 s_W c_W
\biggl(
{ \Pi_{\gamma Z}(M_Z^2)-\Pi_{\gamma Z}(0)
\over M_Z^2}\biggr)
-s_W^2
{ \Pi_{\gamma \gamma}(M_Z^2)\over M_Z^2}\biggr\} \, ,
\label{sdef}
\end{eqnarray}
where $s_W\equiv \sin\theta_W$ and $c_W\equiv \cos\theta_W$
and  any definition of $s_W$ can be used in Eq. \ref{sdef} since
the scheme dependence is higher order.
The contributions to $S,T,$ and $U$ from fourth generation fermions,
squarks, and the scalars of the MSSM
 are given in 
Appendix B\cite{Heinemeyer:2004gx,Altarelli:1990zd,He:2001tp,Hagiwara:1994pw,Peskin:1991sw,Haber:1993wf,Djouadi:2005gj,Dabelstein:1994hb,Drees:1991zk,Drees:1990dx,Cho:1999km}.  Our definition of $U$ differs from that of Ref. 
\cite{He:2001tp} and so should not be compared with those results.
The potential contributions from other MSSM particles such as charginos
and neutralinos decouple for heavy masses\cite{Haber:1993wf} and we omit them here.

Considerable insight can be gained from various limits
of $S,T$ and $U$.  We
begin by considering the contributions from a heavy fermion
generation as defined 
in Eq. \ref{fermdef}\cite{Peskin:1991sw,He:2001tp,Hagiwara:1994pw}.
 The potentially large
isospin violating contributions to $\Delta T_f$ imply that fermions
in an $SU(2)_L$ doublet must have nearly degenerate masses.  For a
fermion doublet with $m_1^2=m_2^2+\delta m_f^2$ and 
$\delta m_f^2 << m_{1,2}^2,M_W^2,M_Z^2$, and $m_{1,2}^2>>M_W^2,M_Z^2$,
\begin{eqnarray}
\Delta S_f&\rightarrow& 
{N_c\over 6 \pi} \biggl\{ 1-2Y_f\biggl({\delta m_f^2\over m_2^2}
\biggr)\biggr\}
\nonumber \\
\Delta T_f& \rightarrow & {N_c\over 48 \pi s_W^2 M_W^2}
{(\delta m_f^2)^2\over  m_2^2}
\nonumber \\
\Delta U_f&\rightarrow & {N_c\over 30 \pi}{(\delta m_f^2)^2\over  m_2^4}\, ,
\end{eqnarray} 
where $N_c=3(1)$ and $Y_f={1\over 6}(-{1\over 2})$ for a quark or lepton
doublet.  
Both $\Delta U_f$ and $\Delta T_f$ are isospin violating, but $\Delta U_f$
is suppressed by a factor of $M_Z^2/m_2^2$ relative to $\Delta T_f$ and
is numerically small.  $\Delta S_f$ does not decouple for large fermion
masses and so poses a potential problem for consistency with
the experimental limits from precision electroweak measurements\cite{gfit,Flacher:2008zq}.
By carefully arranging the $4^{th}$ generation
quark and lepton masses, however,  it is possible find values
of the fermion masses where the contribution to $\Delta S_f$ is 
reduced from its
 value for degenerate fermion partners
of ${N_c\over 6 \pi}$, while still respecting the limits
on $\Delta T_f$\cite{Kribs:2007nz}.  This possibility is due
to the strong correlation between the experimental limits
on $\Delta S$ and $\Delta T$\cite{Kribs:2007nz,Haller:2010zb}.

The $4^{th}$generation squarks and sleptons are denoted by,
\begin{equation}
\left(
\begin{array}{c}
{\tilde t^\prime}_L \\
{\tilde b^\prime}_L
\end{array}
\right)\, ,
\left(
\begin{array}{c}
{\tilde \nu^\prime}_L \\
{\tilde e^\prime}_L
\end{array}
\right)
\, ,
{\tilde t^\prime}_R, {\tilde b^\prime}_R, 
{\tilde e^\prime}_R, {\tilde \nu^\prime}_R
\end{equation}
Consider the limit of no mixing
between the left- and right- handed sfermion
partners, and also no mixing between the sfermion generations.
(The mixing between left- and right- handed sfermions is
included in the formulae in Appendix B).
  In this limit, the the contribution from sfermions
with small mixing between the isospin partners,
${\tilde m}_{t'_L}^2-{\tilde m}_{b'_L}^2<<
{\tilde m}_{t'_L}^2,{\tilde m}_{b'_L}^2$,
 is
\cite{Drees:1990dx,Cho:1999km}, 

\begin{eqnarray}
\Delta S_{sf}& \rightarrow & -{1\over 12 \pi} 
\biggl[3 Y_q
\biggl(
{{\tilde m}_{t'_L}^2-{\tilde m}_{b'_L}^2\over {\tilde m}_{b'_L}^2}
\biggr)
+Y_l\biggl(
{{\tilde m}_{\nu'_L}^2-{\tilde m}_{e'_L}^2\over {\tilde m}_{e'_L}^2}
\biggr)
\biggr]+{\cal O}\biggl({1\over {\tilde m^4}}\biggr)\, .
\end{eqnarray}
(Note that only the scalar partners of the left-handed sfermions contribute
in this limit).
For intermediate values of the sfermion masses, it is possible to arrange
cancellations between the slepton and squark contributions.
In the same limit, the contributions from squarks and sleptons to the
$\Delta T_{sf}$\cite{Haber:1993wf,Heinemeyer:2004gx,Djouadi:2005gj,Cho:1999km},
and $\Delta U_{sf}$\cite{Cho:1999km} parameters 
are\cite{Haber:1993wf,Heinemeyer:2004gx,Djouadi:2005gj,Cho:1999km},
\begin{eqnarray}
\Delta T_{sf}& \rightarrow & 
 {1\over 48 \pi s_W^2 M_W^2}\biggl[
3{({\tilde m_{t'_L}}^2-{\tilde m_{b'_L}}^2)\over
{\tilde m_{b'_L}}^2 }+
{({\tilde m_{\nu'_L}}^2-{\tilde m_{e'_L}}^2)\over 
{\tilde m_{e'_L}}^2}
\biggl]+{\cal O}\biggl({1\over {\tilde m^4}}\biggr)\, .
\nonumber \\
\Delta U_{sf} & \rightarrow & -{1\over 30 \pi}\biggl[
3 
{({\tilde m}_{t'_L}^2-{\tilde m}_{b'_L}^2)^2\over {\tilde m}_{b'_L}^4}
+
{({\tilde m}_{\nu'_L}^2-{\tilde m}_{e'_L}^2)^2\over {\tilde m}_{e'_L}^2}
\biggr]+{\cal O}\biggl({1\over {\tilde m^4}}\biggr)\, .
\end{eqnarray} 
The sfermion contributions decouple for heavy masses
and for sfermions with  $TeV$ scale masses, the effects 
on precision electroweak constraints are small.
In our numerical results, we use the complete amplitudes given in 
Appendix B.  The major effect of heavy sfermions in the 4GMSSM is to increase
the predictions for the neutral Higgs masses,
as shown in Fig. \ref{fg:mh}.

We study the restriction on the 4GMSSM using the fits to 
$\Delta S,~\Delta T$, and $\Delta U$
given by the GFITTER collaboration\cite{gfit}.
\begin{eqnarray}
\Delta S = S-S_{SM}&=& 0.02 \pm 0.11 \nonumber \\
\Delta T = T-T_{SM}&=&0.05\pm 0.12\nonumber \\
\Delta U = U-U_{SM}&=&+ 0.07 \pm 0.12
\label{delts}
\end{eqnarray}
with the Standard Model values defined by $M_{h,ref}=120~GeV$
and $M_t=173.2~GeV$.
The associated correlation matrix is,
\begin{eqnarray}
\rho_{ij}=\left(
\begin{array}{lll}
1.0 & 0.879 & -0.469\nonumber \\
0.879 & 1.0 & -0.716 \nonumber\\
-0.469 & -0.716 & 1.0
 \end{array}
\right)\, .
\end{eqnarray}
$\Delta \chi^2$ is defined as
\begin{equation}
\Delta \chi^2=\Sigma_{ij}(\Delta X_i-\Delta {\hat X}_i)
(\sigma^2)^{-1}_{ij}(\Delta X_i-\Delta {\hat X}_i)\, ,
\end{equation}
where $\Delta {\hat X}_i =\Delta S, \Delta T,$ and
$\Delta U$ are the central values of the fit in Eq.
\ref{delts},  $\Delta X_i=\Delta S, \Delta T$, and
$\Delta U$ 
include the $4^{th}$ generation fermions, sfermions, and
MSSM scalars,
  $\sigma_i$ are the errors
given in Eq. \ref{delts} and $\sigma^2_{ij}=\sigma_i\rho_{ij}\sigma_j$.
The $95\%$ confidence level limit corresponds to $\Delta \chi^2 =
7.815$. 

\begin{figure}[t]
\begin{center}
\includegraphics[scale=0.6]{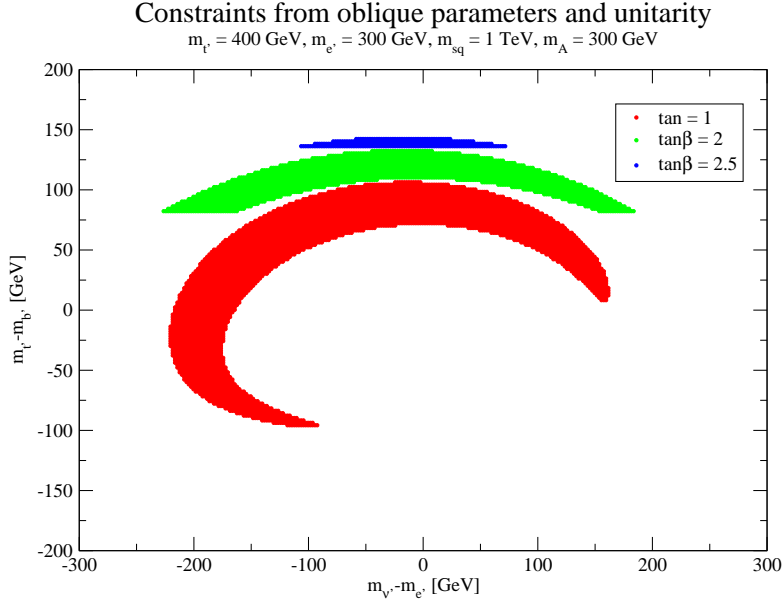}
\caption[]{$95~\%$ confidence level allowed regions from fits
to S, T, and U in the 4GMSSM.
  The requirement of perturbative unitarity is also
imposed.  From top to bottom, the curves correspond to $\tan\beta=2.5, ~2$, 
and $1$.
}
\label{fg:stu1}
\end{center}
\end{figure}

\begin{figure}[t]
\begin{center}
\includegraphics[scale=0.6]{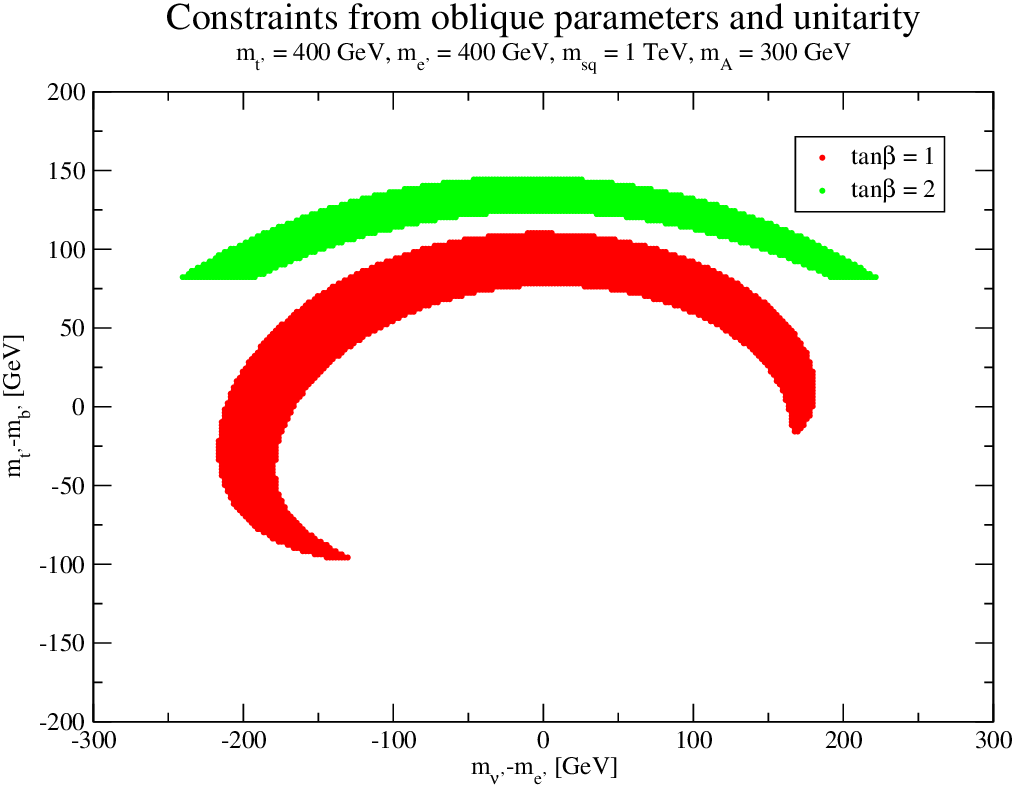}
\caption[]{
$95~\%$ confidence level allowed regions from fits
to S, T, and U in the 4GMSSM.
  The requirement of perturbative unitarity is also
imposed.  From top to bottom, the curves correspond to $\tan\beta=2$ 
and $1$.  The only difference from Fig. 
\ref{fg:stu1} is that $m_{e\prime}=400~GeV$ here.}
\label{fg:stu2}
\end{center}
\end{figure}

\begin{figure}[t]
\begin{center}
\includegraphics[scale=0.6]{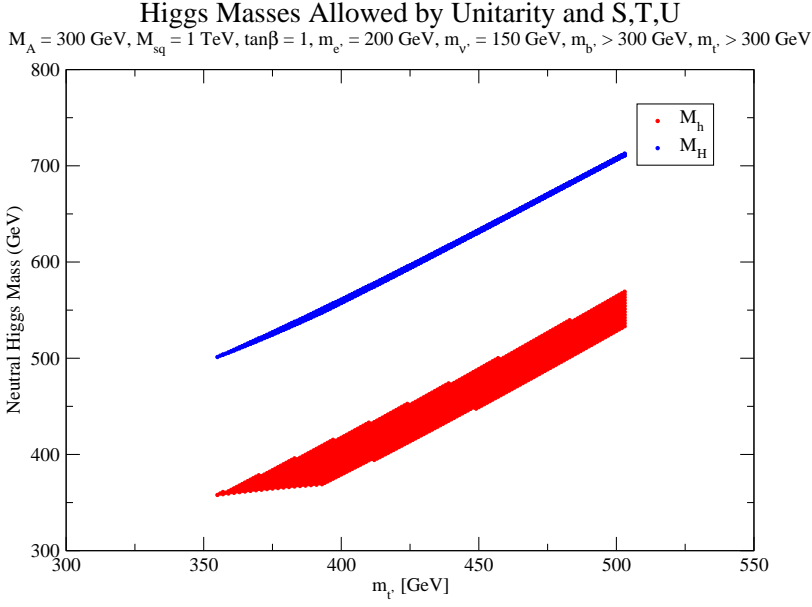}
\caption[]{Neutral Higgs boson masses allowed by precision
electroweak measurements and by unitarity in the 4GMSSM. The
mass difference between $b^\prime$ and $t^\prime$ is scanned over. The
experimental constraints on the $4^{th}$ generation masses
are also imposed.
}
\label{fg:ml}
\end{center}
\end{figure}

In Fig. \ref{fg:stu1}  we show the $95\%$ confidence
level allowed region for $m_{t^\prime}=400~GeV$, 
$M_A=300~GeV$ and $m_{e^{\prime}}=300~GeV$ and
including 4 generations of sfermions with degenerate masses,
 $m_{sq}=1 ~TeV$.
 The difference
between the masses of the quark and lepton  isospin $+{1\over 2}$
and $-{1\over 2}$  doublet
partners  is  scanned over (while imposing the
requirement of perturbative unitarity as
discussed in the previous section) to find the allowed regions. 
We note that the point with all $4^{th}$ generation masses degenerate
is not allowed.  As pointed out in Refs. \cite{Kribs:2007nz,Haller:2010zb,Erler:2010sk}
for the Standard Model case, the 
fermion masses must be carefully tuned to find agreement with 
precision electroweak measurements.
As $\tan\beta$ is increased, the allowed region shrinks and for
the parameters of Figs. \ref{fg:stu1} and \ref{fg:stu2} there is
no allowed region with $\tan\beta > 2.5$.  The Higgs boson masses
vary within these plots according to Eq. \ref{masses}. Fig. \ref{fg:stu2} 
demonstrates the effects of increasing the charged lepton mass.
The effect of increasing the $t^\prime$ mass is shown in Fig. \ref{fg:stu1p}
and we see that the allowed parameter space is significantly shrunk
from Figs. \ref{fg:stu1} and \ref{fg:stu2}.
 In Fig. \ref{fg:ml},
we show the allowed range of Higgs masses corresponding to the
scan of Fig.  \ref{fg:stu1} and imposing
the experimental constraints on $4^{th}$
generation masses.  It is apparent that the 4GMSSM requires
highly tuned fermion masses in order to be viable.

\begin{figure}[t]
\begin{center}
\includegraphics[scale=0.6]{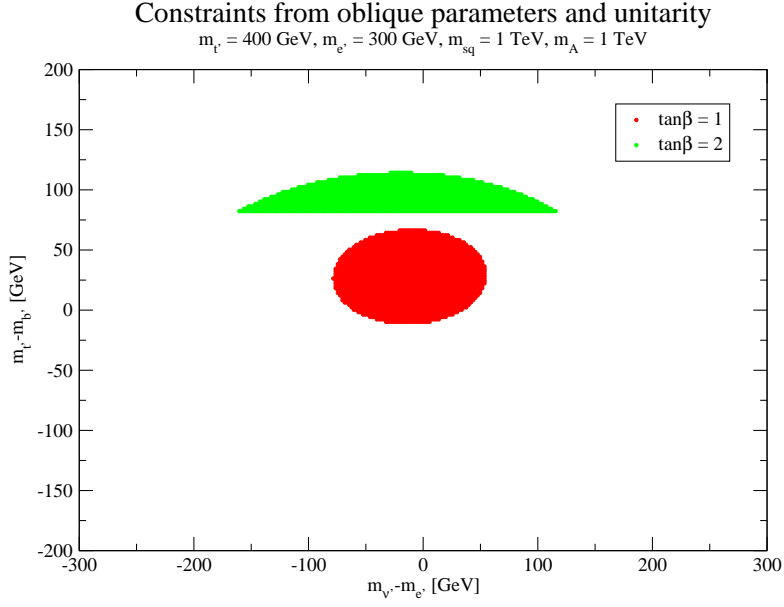}
\caption[]{
$95~\%$ confidence level allowed regions from fits
to S, T, and U in the 4GMSSM.
  The requirement of perturbative unitarity is also
imposed.  From top to bottom, the curves correspond to $\tan\beta=2$ 
and $1$.  The only difference from Fig. 
\ref{fg:stu1} is that $m_A=1~TeV$ here.}
\label{fg:stu5}
\end{center}
\end{figure}

\begin{figure}[t]
\begin{center}
\includegraphics[scale=0.6]{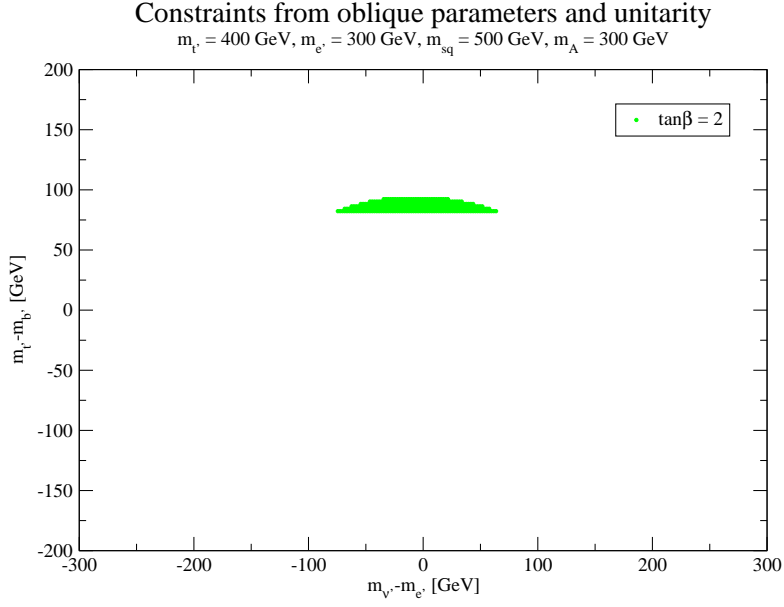}
\caption[]{$95~\%$ confidence level allowed regions from fits
to S, T, and U in the 4GMSSM.
  The requirement of perturbative unitarity is also
imposed.  The only difference from Fig. 
\ref{fg:stu1} is that $m_sq=500~GeV$ here.}
\label{fg:stu3}
\end{center}
\end{figure}

The effect of increasing $m_A$ (and hence $M_h$)
is shown in Fig. \ref{fg:stu5} and we see only a very small
region of allowed parameters.  
In Fig. \ref{fg:stu3}, we show the effects of lowering the
sfermion masses to $500~GeV$ and see that the allowed region 
shrinks considerably.  This is due not to the effects of sfermion
contributions to the electroweak limits, but rather to the
change in Higgs mass corresponding to the heavier  squark masses.

\begin{figure}[t]
\begin{center}
\includegraphics[scale=0.6]{ma300mt500.eps}
\caption[]{
$95~\%$ confidence level allowed regions from fits
to S, T, and U in the 4GMSSM.
  The requirement of perturbative unitarity is also
imposed.  From top to bottom, the curves correspond to $\tan\beta=2$ 
and $1$.  The only difference from Fig. 
\ref{fg:stu1} is that $m_{t\prime}=500~GeV$ here.}
\label{fg:stu1p}
\end{center}
\end{figure}

\section{Conclusions}
\label{conc}
We have studied an extension of the MSSM with $4$ generations of chiral
fermions.  The existence of a $4^{th}$ generation allows the lightest
neutral Higgs boson mass to be considerably heavier than in the Standard
Model.  However, imposing the restrictions of perturbative unitarity 
and constraints from precision electroweak measurements requires
$\tan\beta\sim 1$ and extremely fine-tuned values of the fermion
masses.

\section*{Acknowledgements}
This work is supported by the United States Department of Energy under
Grant DE-AC02-98CH10886.

\section*{Appendix A: Neutral Higgs Mass}
The neutral Higgs masses are found from the eigenvectors of the
matrix\cite{Gunion:1989we}:
\begin{equation}
M^2=
\left(
\begin{array}{cc} 
M_{11}& M_{21}\\
M_{12}& M_{22}\end{array}
\right)
\end{equation}
where $M_{ij}\equiv M_{ij,tree}+\Delta_{ij}$
The tree level values are (where $c_\beta=cos\beta$
and $s_\beta=sin\beta$),
\begin{eqnarray}
M_{11,tree}&=&M_A^2s_\beta^2+M_Z^2c_\beta^2\nonumber \\
M_{12,tree}&=&-(M_A^2+M_Z^2)s_\beta c_\beta\\
M_{22,tree}&=& M_A^2 c_\beta^2 +M_Z^2 s_\beta^2
\end{eqnarray}
At one-loop the effects of a heavy $4^{th}$ generation
on the neutral Higgs masses,
including only the  leading logarithms and assuming no mixing
in the sfermion sector,  are \cite{Haber:1990aw,Ellis:1990nz}
\begin{eqnarray}
\Delta_{11}&=& 
{\hat \epsilon_b}
=\Sigma_{i=b^\prime,e^\prime}
{N_{c}G_F\over 2\sqrt{2} \pi^2 }{m_i^4\over c_\beta^2}
\ln\biggl(
{{\tilde m_{i1}}^2{\tilde m_{i2}}^2\over m_i^4}\biggr)
\nonumber \\
\Delta_{22}
&=& {\hat \epsilon_t}=
\Sigma_{i=t^\prime,\nu^\prime}{N_{c}G_F\over 2\sqrt{2} \pi^2 }
{m_i^4\over s_\beta^2}
\ln\biggl(
{{\tilde m_{i1}}^2{\tilde m_{i2}}^2\over m_i^4}\biggr)
\nonumber \\
\Delta_{12}&=& 0\, ,
\end{eqnarray}
where ${\tilde m_{i1}}$ and ${\tilde m_{i2}}$ are the physical sfermion
masses associated with $f_i$.

The neutral  Higgs boson  masses are then,
\begin{eqnarray}
m_{H,h}^2&=&
{1\over 2}
\biggl\{
M_A^2+M_Z^2+{\hat \epsilon_b}+{\hat \epsilon_t}
\pm\biggl[ 
(M_A^2+M_Z^2)^2
-4c_{2\beta}^2 M_A^2 M_Z^2
\nonumber \\ &&
+({\hat\epsilon_b}-{\hat \epsilon_t})\biggl( 2c_{2\beta}(M_Z^2-M_A^2)+
{\hat\epsilon_b}-{\hat\epsilon_t}\biggr)
\biggr]^{1/2}\biggr\}\, .
\label{masses}
\end{eqnarray}
Formulae including non-logarithmic terms and $2$- loop  contributions
to the neutral Higgs boson masses
can be found in Ref. \cite{Djouadi:2005gj}. 
 However, since we assume
very heavy $4^{th}$ generation quarks, we expect Eq. \ref{masses} to
be a  good approximation.

The mixing angle (which we use to define the fermion and sfermion couplings)
is
\begin{equation}
sin 2\alpha =
{2M_{11}^2\over \sqrt{(M_{11}^2-M_{22}^2)^2+4M_{12}^4}}
\end{equation}

\section*{Appendix B: Summary  of S,T,U formula}

From a heavy $SU(2)$ Standard
Model like doublet of fermions with masses $(m_1,m_2)$ and
isospin $Y_f$\cite{Kniehl:1990mq,Hagiwara:1994pw,He:2001tp},
\begin{eqnarray}
\Delta S_f &=& {N_c\over 6 \pi M_Z^2}\biggl\{ 2 Y_f\biggl[
- M_Z^2\ln\biggl({m_1^2\over m_2^2}\biggr)
+(2 m_1^2+M_Z^2)f_0(m_Z^2,m_1^2)
\nonumber \\
&& -(2 m_2^2+M_Z^2)f_0(m_Z^2,m_2^2)\biggr] 
+3 m_1^2 f_0(m_Z^2,m_1^2)
+3 m_2^2 f_0(m_Z^2,m_2^2)\biggr\}
\nonumber \\
\Delta T_f&=& {N_c\over 16 \pi s_W^2 M_W^2}F(m_1^2,m_2^2)
\nonumber \\
\Delta U_f&=& -{N_c\over \pi M_W^2}
\biggl\{
\biggl[{M_W^2\over 3}-{m_1^2+m_2^2\over 6}-
{(m_1^2-m_2^2)^2 \over 6M_W^2}\biggr]f(M_W^2,m_1^2,m_2^2)
+{m_1^2+m_2^2\over 4}
\nonumber \\ &&
+{(m_1^2-m_2^2)^2\over 6M_W^2}
-\biggl[{m_1^4-m_2^4\over 12M_W^2}
+{m_1^2 m_2^2\over 2(m_1^2-m_2^2)}+{m_1^2-m_2^2\over 12}\biggr] 
\ln\biggl({m_1^2\over m_2^2}\biggr)
\nonumber \\ &&
+c_W^2\biggl[{m_1^2-M_Z^2 \over 6}f_0(M_Z^2,m_1^2)
+{m_2^2-M_Z^2 \over 6}f_0(M_Z^2,m_2^2)\biggr]
\biggl\}
\end{eqnarray}
where
\begin{eqnarray}
F(x,y)&=&x+y-{2xy\over x-y}\log\biggl({x\over y}\biggr)
\nonumber \\
f(p^2,m_1^2,m_2^2)
&=&-\int_0^1 dx ~ \log\biggl[x{m_2\over m_1}
+(1-x){m_1\over m_2}-x(1-x){p^2\over m_1 m_2}\biggr]
\nonumber \\
f_0(p^2,m^2)&\equiv& f(p^2,m^2,m^2) \nonumber \\
&=&2-2 \sqrt{{4m^2\over p^2}-1} \tan ^{-1}\biggl(
{1\over \sqrt{{4m^2\over p^2}-1}}\biggr)\, .
\end{eqnarray}
$N_c=3$ for quarks and $1$ for a lepton
doublet. The hypercharge is $Y_q={1\over 6}$ 
for a quark doublet and $Y_l=-{1\over 2}$
for a lepton doublet.

We define ${\tilde m}_{t1},{\tilde m}_{t2},
{\tilde m}_{b1}$, and ${\tilde m}_{b1}$
to be the physical masses of the $4^{th}$ generation squarks with the
mixing angles, $\theta_t$ and $\theta_b$, ($s_b\equiv sin\theta_b$, etc).
From the $4^{th}$ generation squarks\cite{Haber:1993wf,Cho:1999km}:
\begin{eqnarray}
\Delta S_{sf} &=& -{N_c\over 2 \pi M_Z^2}\biggl\{
Q_t\biggl[c_t^2 F_5(M_Z^2,{\tilde m}_{t1}^2,{\tilde m}_{t1}^2)
+s_t^2 F_5(M_Z^2,{\tilde m}_{t2}^2,{\tilde m}_{t2}^2)\biggr]
\nonumber \\ &&
-Q_b\biggl[c_b^2 F_5(M_Z^2,{\tilde m}_{b1}^2,{\tilde m}_{b1}^2)
+s_b^2 F_5(M_Z^2,{\tilde m}_{b2}^2,{\tilde m}_{b2}^2)\biggr]
\nonumber \\
&&-{1\over 2} \biggl[ c_t^4 F_5(M_Z^2,{\tilde m}_{t1}^2,{\tilde m}_{t1}^2)
+2c_t^2 s_t^2 {\overline F}_5(M_Z^2,{\tilde m}_{t1}^2,{\tilde m}_{t2}^2)
\nonumber \\
&&+ s_t^4 F_5(M_Z^2,{\tilde m}_{t2}^2,{\tilde m}_{t2}^2)
+ c_b^4 F_5(M_Z^2,{\tilde m}_{b1}^2,{\tilde m}_{b1}^2)
\nonumber \\ &&
+2c_b^2 s_b^2 {\overline F}_5(M_Z^2,{\tilde m}_{b1}^2,{\tilde m}_{b2}^2)
+ s_b^4 F_5(M_Z^2,{\tilde m}_{b2}^2,{\tilde m}_{b2}^2)\biggr]
\biggr\}\nonumber \\
\Delta T_{sf}&=&{N_c\over 16 \pi M_W^2 s_W^2}
\biggl\{
 -s_t^2 c_t^2 F({\tilde m}_{t1}^2,{\tilde m}_{t2}^2)
-s_b^2 c_b^2F({\tilde m}_{b1}^2,{\tilde m}_{b2}^2)
\nonumber \\
&& + c_t^2c_b^2 F({\tilde m}_{t1}^2,{\tilde m}_{b1}^2)
+c_t^2 s_b^2 F({\tilde m}_{t1}^2,{\tilde m}_{b2}^2)
\nonumber \\
&& + s_t^2c_b^2 F({\tilde m}_{t2}^2,{\tilde m}_{b1}^2)
+s_t^2 s_b^2 F({\tilde m}_{t2}^2,{\tilde m}_{b2}^2)
\biggr\}\nonumber \\
\Delta U_{sf} &=&
-{N_c\over 4 \pi M_W^2}\biggl\{
c_W^2\bigg(c_t^4 F_5(M_Z^2,{\tilde m}_{t1}^2,{\tilde m}_{t1}^2)
+s_t^4 F_5(M_Z^2,{\tilde m}_{t2}^2,{\tilde m}_{t2}^2)
\nonumber \\ &&
+2 s_t^2 c_t^2
{\overline F}_5(M_Z^2,{\tilde m}_{t1}^2,{\tilde m}_{t2}^2)
+c_b^4 F_5(M_Z^2,{\tilde m}_{b1}^2,{\tilde m}_{b1}^2)
\nonumber \\ &&
+s_b^4 F_5(M_Z^2,{\tilde m}_{b2}^2,{\tilde m}_{b2}^2)
+2 s_b^2 c_b^2
{\overline F}_5(M_Z^2,{\tilde m}_{b1}^2,{\tilde m}_{b2}^2)
\biggr)
\nonumber \\ &&
-2\biggl(
c_t^2 c_b^2 {\overline F}_5(M_W^2,{\tilde m}_{t1}^2,{\tilde m}_{b1}^2)
+ s_t^2c_b^2 
{\overline F}_5(M_W^2,{\tilde m}_{t2}^2,{\tilde m}_{b1}^2)
\nonumber \\ &&
+ s_b^2c_t^2 
{\overline F}_5(M_W^2,{\tilde m}_{t1}^2,{\tilde m}_{b2}^2)
+s_t^2 s_b^2 {\overline F}_5(M_Z^2,{\tilde m}_{t2}^2,{\tilde m}_{b2}^2)
\biggr)
\biggr\}
\end{eqnarray}
where,
\begin{eqnarray}
F_5(p^2,m_1^2,m_2^2)
&=&\int_0^1 dx ~ 
\biggl[(1-2x)(m_1^2-m_2^2)+(1-2x)^2 p^2\biggr]\Lambda
\nonumber \\
{\overline F}_5(p^2,m_1^2,m_2^2)
&=& F_5(p^2,m_1^2,m_2^2)-F_5(0,m_1^2,m_2^2)\nonumber \\
\Lambda&=&\log\biggl((1-x)m_1^2+xm_2^2-x(1-x)p^2\biggl)
\end{eqnarray}
The contributions from the $4^{th}$ generation sleptons are
found in an analogous manner.

From Higgs scalars, (where $M_{h,ref}$ is the value of the
 Standard Model
Higgs boson mass for which the fits are performed.), the
contribution of the MSSM scalars is\cite{Haber:1993wf,Haber:1999zh},
\begin{eqnarray}
\Delta S_H&=&S_H(M_h,M_H,M_A,M_{H\pm})-S_{sm}(M_{h,ref})\nonumber \\
&=& {1\over 4\pi} 
\biggl\{
\sin^2(\beta-\alpha) \biggl[{1\over M_Z^2}{\overline F}_{5}(M_Z^2,
M_H^2,M_A^2)
+F_2(M_Z^2, M_Z^2,M_h^2)\biggr]\nonumber \\ &&
+\cos^2(\beta-\alpha) \biggl[{1\over M_Z^2}{\overline F}_{5}(M_Z^2,
M_h^2,M_A^2)
+F_2(M_Z^2, M_Z^2,M_H^2)\biggr]\nonumber \\
&&-{1\over M_Z^2}{\overline F}_{5}(M_Z^2,M_{H\pm}^2,M_{H\pm}^2)
-F_2(M_Z^2,M_Z^2,M_{h,ref}^2)\biggr\}
\nonumber \\
&&
\nonumber \\
\Delta T_H &=&T_H(M_h,M_H,M_A,M_{H\pm})-T_{sm}(M_{h,ref})\nonumber \\
&=&{1\over 32 \pi M_W^2 s_W^2}
\biggl\{\cos^2(\beta-\alpha)\biggl[
G_3(M_W^2, M_Z^2, M_H^2)
+G_3(M_{H\pm}^2,M_A^2,M_h^2)\biggr]
\nonumber \\
&& +\sin^2(\beta-\alpha) \biggl[
G_3(M_W^2, M_Z^2, M_h^2)
+G_3(M_{H\pm}^2,M_A^2,M_H^2)
\nonumber \\ &&
-8M_Z^2 G_4(M_Z^2,M_H^2,M_h^2)
+8M_W^2 G_4(M_W^2,M_H^2,M_h^2)\biggr]
\nonumber \\ &&
+ F(M_{H\pm}^2, M_A^2)-G_3(M_W^2,M_Z^2, M_{h,ref}^2)
\nonumber \\ && 
 +8M_Z^2 G_4(M_Z^2,M_H^2,M_{h,ref}^2)
-8M_W^2 G_4(M_W^2,M_H^2,M_{h,ref}^2)
\biggr\}\nonumber \\
\Delta U_H&=& 
U_H(M_h,M_H,M_A,M_{H\pm})
-U_{sm}(M_{h,ref})\nonumber \\
&=&{1\over 4\pi}\biggl\{
-{1\over M_Z^2}{\overline F}_{5}(M_Z^2,M_{H_\pm}^2,M_{H^\pm}^2)
+{1\over M_W^2}{\overline F}_{5}(M_W^2,M_{A}^2,M_{H^\pm}^2)
\nonumber \\ &&
+\sin^2(\beta-\alpha) \biggl[
{1\over M_W^2}{\overline F}_{5}(M_W^2,M_H^2,M_{H^\pm}^2)
-{1\over M_Z^2}{\overline F}_{5}(M_Z^2,M_H^2,M_A^2)
\nonumber \\ &&
+F_2(M_W^2, M_W^2,M_h^2)
-F_2(M_Z^2, M_Z^2,M_h^2)\biggr]
+\cos^2(\beta-\alpha) \biggl[
-{1\over M_Z^2}{\overline F}_{5}(M_Z^2,M_h^2,M_A^2)
\nonumber \\ &&
+{1\over M_W^2}{\overline F}_{5}(M_W^2,M_{H_\pm}^2,M_h^2)
+F_2(M_W^2, M_W^2,M_H^2)-
F_2(M_Z^2, M_Z^2,M_H^2)\biggr]\nonumber \\ &&
-F_2(M_W^2,M_W^2,M_{h,ref}^2)
+ F_2(M_Z^2, M_Z^2,M_{h,ref}^2)\biggr]\biggr\}
\end{eqnarray}
where
\begin{eqnarray}
{\hat B}_0(p^2,m_1^2,m_2^2)
&=&B_0(p^2,m_1^2,m_2^2)-B_0(0,m_1^2,m_2^2)
\nonumber \\
B_0(p^2,m_1^2,m_2^2&=&{1\over \epsilon}\biggl({4\pi\mu^2\over m_1
m_2}\biggr)^\epsilon\Gamma(1+\epsilon)+f(p^2,m_1^2,m_2^2)
\nonumber \\
F_2(p^2,m_1^2,m_2^2)&=& {1\over p^2}{\overline F}_{5}(p^2,m_1^2,m_2^2)
-4 {\hat B}_0(p^2,m_1^2,m_2^2)
\nonumber \\
G_3(m_1^2,m_2^2,m_3^2)&=&F(m_1^2,m_3^2)-F(m_2^2,m_3^2)
\nonumber \\
G_4(m_1^2,m_2^2,m_3^2)&=&m_3^2\log\biggl({m_1^2\over m_3^2}\biggr)
-m_2^2\log\biggl({m_1^2\over m_2^2}\biggr)
\end{eqnarray}
\bibliography{g4}

\begin{thebibliography}{47}
\expandafter\ifx\csname natexlab\endcsname\relax\def\natexlab#1{#1}\fi
\expandafter\ifx\csname bibnamefont\endcsname\relax
  \def\bibnamefont#1{#1}\fi
\expandafter\ifx\csname bibfnamefont\endcsname\relax
  \def\bibfnamefont#1{#1}\fi
\expandafter\ifx\csname citenamefont\endcsname\relax
  \def\citenamefont#1{#1}\fi
\expandafter\ifx\csname url\endcsname\relax
  \def\url#1{\texttt{#1}}\fi
\expandafter\ifx\csname urlprefix\endcsname\relax\def\urlprefix{URL }\fi
\providecommand{\bibinfo}[2]{#2}
\providecommand{\eprint}[2][]{\url{#2}}

\bibitem[{\citenamefont{Holdom}(2007)}]{Holdom:2007nw}
\bibinfo{author}{\bibfnamefont{B.}~\bibnamefont{Holdom}},
  \bibinfo{journal}{JHEP} \textbf{\bibinfo{volume}{03}}, \bibinfo{pages}{063}
  (\bibinfo{year}{2007}), \eprint{hep-ph/0702037}.

\bibitem[{\citenamefont{Soni et~al.}(2010)\citenamefont{Soni, Alok, Giri,
  Mohanta, and Nandi}}]{Soni:2010xh}
\bibinfo{author}{\bibfnamefont{A.}~\bibnamefont{Soni}},
  \bibinfo{author}{\bibfnamefont{A.~K.} \bibnamefont{Alok}},
  \bibinfo{author}{\bibfnamefont{A.}~\bibnamefont{Giri}},
  \bibinfo{author}{\bibfnamefont{R.}~\bibnamefont{Mohanta}}, \bibnamefont{and}
  \bibinfo{author}{\bibfnamefont{S.}~\bibnamefont{Nandi}},
  \bibinfo{journal}{Phys. Rev.} \textbf{\bibinfo{volume}{D82}},
  \bibinfo{pages}{033009} (\bibinfo{year}{2010}), \eprint{1002.0595}.

\bibitem[{\citenamefont{Aaltonen et~al.}(2010{\natexlab{a}})}]{Aaltonen:2009nr}
\bibinfo{author}{\bibfnamefont{T.}~\bibnamefont{Aaltonen}} \bibnamefont{et~al.}
  (\bibinfo{collaboration}{CDF}), \bibinfo{journal}{Phys. Rev. Lett.}
  \textbf{\bibinfo{volume}{104}}, \bibinfo{pages}{091801}
  (\bibinfo{year}{2010}{\natexlab{a}}), \eprint{0912.1057}.

\bibitem[{\citenamefont{Lister}(2008)}]{Lister:2008is}
\bibinfo{author}{\bibfnamefont{A.}~\bibnamefont{Lister}}
  (\bibinfo{collaboration}{CDF}) (\bibinfo{year}{2008}), \eprint{0810.3349}.

\bibitem[{\citenamefont{Flacco et~al.}(2010)\citenamefont{Flacco, Whiteson,
  Tait, and Bar-Shalom}}]{Flacco:2010rg}
\bibinfo{author}{\bibfnamefont{C.~J.} \bibnamefont{Flacco}},
  \bibinfo{author}{\bibfnamefont{D.}~\bibnamefont{Whiteson}},
  \bibinfo{author}{\bibfnamefont{T.~M.~P.} \bibnamefont{Tait}},
  \bibnamefont{and}
  \bibinfo{author}{\bibfnamefont{S.}~\bibnamefont{Bar-Shalom}}
  (\bibinfo{year}{2010}), \eprint{1005.1077}.

\bibitem[{\citenamefont{Kribs et~al.}(2007)\citenamefont{Kribs, Plehn,
  Spannowsky, and Tait}}]{Kribs:2007nz}
\bibinfo{author}{\bibfnamefont{G.~D.} \bibnamefont{Kribs}},
  \bibinfo{author}{\bibfnamefont{T.}~\bibnamefont{Plehn}},
  \bibinfo{author}{\bibfnamefont{M.}~\bibnamefont{Spannowsky}},
  \bibnamefont{and} \bibinfo{author}{\bibfnamefont{T.~M.~P.}
  \bibnamefont{Tait}}, \bibinfo{journal}{Phys. Rev.}
  \textbf{\bibinfo{volume}{D76}}, \bibinfo{pages}{075016}
  (\bibinfo{year}{2007}), \eprint{0706.3718}.

\bibitem[{\citenamefont{Hung and Sher}(2008)}]{Hung:2007ak}
\bibinfo{author}{\bibfnamefont{P.~Q.} \bibnamefont{Hung}} \bibnamefont{and}
  \bibinfo{author}{\bibfnamefont{M.}~\bibnamefont{Sher}},
  \bibinfo{journal}{Phys. Rev.} \textbf{\bibinfo{volume}{D77}},
  \bibinfo{pages}{037302} (\bibinfo{year}{2008}), \eprint{0711.4353}.

\bibitem[{\citenamefont{Eberhardt et~al.}(2010)\citenamefont{Eberhardt, Lenz,
  and Rohrwild}}]{Eberhardt:2010bm}
\bibinfo{author}{\bibfnamefont{O.}~\bibnamefont{Eberhardt}},
  \bibinfo{author}{\bibfnamefont{A.}~\bibnamefont{Lenz}}, \bibnamefont{and}
  \bibinfo{author}{\bibfnamefont{J.}~\bibnamefont{Rohrwild}}
  (\bibinfo{year}{2010}), \eprint{1005.3505}.

\bibitem[{\citenamefont{Bobrowski et~al.}(2009)\citenamefont{Bobrowski, Lenz,
  Riedl, and Rohrwild}}]{Bobrowski:2009ng}
\bibinfo{author}{\bibfnamefont{M.}~\bibnamefont{Bobrowski}},
  \bibinfo{author}{\bibfnamefont{A.}~\bibnamefont{Lenz}},
  \bibinfo{author}{\bibfnamefont{J.}~\bibnamefont{Riedl}}, \bibnamefont{and}
  \bibinfo{author}{\bibfnamefont{J.}~\bibnamefont{Rohrwild}},
  \bibinfo{journal}{Phys. Rev.} \textbf{\bibinfo{volume}{D79}},
  \bibinfo{pages}{113006} (\bibinfo{year}{2009}), \eprint{0902.4883}.

\bibitem[{\citenamefont{Chanowitz}(2010)}]{Chanowitz:2010bm}
\bibinfo{author}{\bibfnamefont{M.~S.} \bibnamefont{Chanowitz}}
  (\bibinfo{year}{2010}), \eprint{1007.0043}.

\bibitem[{\citenamefont{Haller and Collaboration}(2010)}]{Haller:2010zb}
\bibinfo{author}{\bibfnamefont{J.}~\bibnamefont{Haller}} \bibnamefont{and}
  \bibinfo{author}{\bibfnamefont{G.}~\bibnamefont{Collaboration}}
  (\bibinfo{year}{2010}), \eprint{1006.0003}.

\bibitem[{\citenamefont{Flacher et~al.}(2009)}]{Flacher:2008zq}
\bibinfo{author}{\bibfnamefont{H.}~\bibnamefont{Flacher}} \bibnamefont{et~al.},
  \bibinfo{journal}{Eur. Phys. J.} \textbf{\bibinfo{volume}{C60}},
  \bibinfo{pages}{543} (\bibinfo{year}{2009}), \eprint{0811.0009}.

\bibitem[{\citenamefont{Collaboration}()}]{gfit}
\bibinfo{author}{\bibfnamefont{G.}~\bibnamefont{Collaboration}},
  \bibinfo{note}{http://gfitter.desy.de/}.

\bibitem[{\citenamefont{Anastasiou et~al.}(2010)\citenamefont{Anastasiou,
  Boughezal, and Furlan}}]{Anastasiou:2010bt}
\bibinfo{author}{\bibfnamefont{C.}~\bibnamefont{Anastasiou}},
  \bibinfo{author}{\bibfnamefont{R.}~\bibnamefont{Boughezal}},
  \bibnamefont{and} \bibinfo{author}{\bibfnamefont{E.}~\bibnamefont{Furlan}},
  \bibinfo{journal}{JHEP} \textbf{\bibinfo{volume}{06}}, \bibinfo{pages}{101}
  (\bibinfo{year}{2010}), \eprint{1003.4677}.

\bibitem[{\citenamefont{Aaltonen et~al.}(2010{\natexlab{b}})}]{Aaltonen:2010sv}
\bibinfo{author}{\bibfnamefont{T.}~\bibnamefont{Aaltonen}} \bibnamefont{et~al.}
  (\bibinfo{collaboration}{CDF}) (\bibinfo{year}{2010}{\natexlab{b}}),
  \eprint{1005.3216}.

\bibitem[{\citenamefont{Litsey and Sher}(2009)}]{Litsey:2009rp}
\bibinfo{author}{\bibfnamefont{S.}~\bibnamefont{Litsey}} \bibnamefont{and}
  \bibinfo{author}{\bibfnamefont{M.}~\bibnamefont{Sher}},
  \bibinfo{journal}{Phys. Rev.} \textbf{\bibinfo{volume}{D80}},
  \bibinfo{pages}{057701} (\bibinfo{year}{2009}), \eprint{0908.0502}.

\bibitem[{\citenamefont{Hou}(2009)}]{Hou:2008xd}
\bibinfo{author}{\bibfnamefont{W.-S.} \bibnamefont{Hou}},
  \bibinfo{journal}{Chin. J. Phys.} \textbf{\bibinfo{volume}{47}},
  \bibinfo{pages}{134} (\bibinfo{year}{2009}), \eprint{0803.1234}.

\bibitem[{\citenamefont{Kikukawa et~al.}(2009)\citenamefont{Kikukawa, Kohda,
  and Yasuda}}]{Kikukawa:2009mu}
\bibinfo{author}{\bibfnamefont{Y.}~\bibnamefont{Kikukawa}},
  \bibinfo{author}{\bibfnamefont{M.}~\bibnamefont{Kohda}}, \bibnamefont{and}
  \bibinfo{author}{\bibfnamefont{J.}~\bibnamefont{Yasuda}},
  \bibinfo{journal}{Prog. Theor. Phys.} \textbf{\bibinfo{volume}{122}},
  \bibinfo{pages}{401} (\bibinfo{year}{2009}), \eprint{0901.1962}.

\bibitem[{\citenamefont{Fok and Kribs}(2008)}]{Fok:2008yg}
\bibinfo{author}{\bibfnamefont{R.}~\bibnamefont{Fok}} \bibnamefont{and}
  \bibinfo{author}{\bibfnamefont{G.~D.} \bibnamefont{Kribs}},
  \bibinfo{journal}{Phys. Rev.} \textbf{\bibinfo{volume}{D78}},
  \bibinfo{pages}{075023} (\bibinfo{year}{2008}), \eprint{0803.4207}.

\bibitem[{\citenamefont{Chanowitz et~al.}(1979)\citenamefont{Chanowitz, Furman,
  and Hinchliffe}}]{Chanowitz:1978mv}
\bibinfo{author}{\bibfnamefont{M.~S.} \bibnamefont{Chanowitz}},
  \bibinfo{author}{\bibfnamefont{M.~A.} \bibnamefont{Furman}},
  \bibnamefont{and}
  \bibinfo{author}{\bibfnamefont{I.}~\bibnamefont{Hinchliffe}},
  \bibinfo{journal}{Nucl. Phys.} \textbf{\bibinfo{volume}{B153}},
  \bibinfo{pages}{402} (\bibinfo{year}{1979}).

\bibitem[{\citenamefont{Godbole et~al.}(2010)\citenamefont{Godbole, Vempati,
  and Wingerter}}]{Godbole:2009sy}
\bibinfo{author}{\bibfnamefont{R.~M.} \bibnamefont{Godbole}},
  \bibinfo{author}{\bibfnamefont{S.~K.} \bibnamefont{Vempati}},
  \bibnamefont{and}
  \bibinfo{author}{\bibfnamefont{A.}~\bibnamefont{Wingerter}},
  \bibinfo{journal}{JHEP} \textbf{\bibinfo{volume}{03}}, \bibinfo{pages}{023}
  (\bibinfo{year}{2010}), \eprint{0911.1882}.

\bibitem[{\citenamefont{Murdock et~al.}(2008)\citenamefont{Murdock, Nandi, and
  Tavartkiladze}}]{Murdock:2008rx}
\bibinfo{author}{\bibfnamefont{Z.}~\bibnamefont{Murdock}},
  \bibinfo{author}{\bibfnamefont{S.}~\bibnamefont{Nandi}}, \bibnamefont{and}
  \bibinfo{author}{\bibfnamefont{Z.}~\bibnamefont{Tavartkiladze}},
  \bibinfo{journal}{Phys. Lett.} \textbf{\bibinfo{volume}{B668}},
  \bibinfo{pages}{303} (\bibinfo{year}{2008}), \eprint{0806.2064}.

\bibitem[{\citenamefont{Gunion et~al.}(1996)\citenamefont{Gunion, McKay, and
  Pois}}]{Gunion:1995tp}
\bibinfo{author}{\bibfnamefont{J.~F.} \bibnamefont{Gunion}},
  \bibinfo{author}{\bibfnamefont{D.~W.} \bibnamefont{McKay}}, \bibnamefont{and}
  \bibinfo{author}{\bibfnamefont{H.}~\bibnamefont{Pois}},
  \bibinfo{journal}{Phys. Rev.} \textbf{\bibinfo{volume}{D53}},
  \bibinfo{pages}{1616} (\bibinfo{year}{1996}), \eprint{hep-ph/9507323}.

\bibitem[{\citenamefont{Hashimoto}(2010)}]{Hashimoto:2010at}
\bibinfo{author}{\bibfnamefont{M.}~\bibnamefont{Hashimoto}},
  \bibinfo{journal}{Phys. Rev.} \textbf{\bibinfo{volume}{D81}},
  \bibinfo{pages}{075023} (\bibinfo{year}{2010}), \eprint{1001.4335}.

\bibitem[{\citenamefont{Carena et~al.}(1996)\citenamefont{Carena, Haber, and
  Wagner}}]{Carena:1995ep}
\bibinfo{author}{\bibfnamefont{M.~S.} \bibnamefont{Carena}},
  \bibinfo{author}{\bibfnamefont{H.~E.} \bibnamefont{Haber}}, \bibnamefont{and}
  \bibinfo{author}{\bibfnamefont{C.~E.~M.} \bibnamefont{Wagner}},
  \bibinfo{journal}{Nucl. Phys.} \textbf{\bibinfo{volume}{B472}},
  \bibinfo{pages}{55} (\bibinfo{year}{1996}), \eprint{hep-ph/9512446}.

\bibitem[{\citenamefont{Gunion et~al.}(1989)\citenamefont{Gunion, Haber, Kane,
  and Dawson}}]{Gunion:1989we}
\bibinfo{author}{\bibfnamefont{J.~F.} \bibnamefont{Gunion}},
  \bibinfo{author}{\bibfnamefont{H.~E.} \bibnamefont{Haber}},
  \bibinfo{author}{\bibfnamefont{G.~L.} \bibnamefont{Kane}}, \bibnamefont{and}
  \bibinfo{author}{\bibfnamefont{S.}~\bibnamefont{Dawson}},
  \emph{\bibinfo{title}{{THE HIGGS HUNTER'S GUIDE}}}
  (\bibinfo{publisher}{Adison-Wesley}, \bibinfo{year}{1989}).

\bibitem[{\citenamefont{De~Pree et~al.}(2009)\citenamefont{De~Pree, Marshall,
  and Sher}}]{DePree:2009ed}
\bibinfo{author}{\bibfnamefont{E.}~\bibnamefont{De~Pree}},
  \bibinfo{author}{\bibfnamefont{G.}~\bibnamefont{Marshall}}, \bibnamefont{and}
  \bibinfo{author}{\bibfnamefont{M.}~\bibnamefont{Sher}},
  \bibinfo{journal}{Phys. Rev.} \textbf{\bibinfo{volume}{D80}},
  \bibinfo{pages}{037301} (\bibinfo{year}{2009}), \eprint{0906.4500}.

\bibitem[{\citenamefont{Ellis et~al.}(1991)\citenamefont{Ellis, Ridolfi, and
  Zwirner}}]{Ellis:1990nz}
\bibinfo{author}{\bibfnamefont{J.~R.} \bibnamefont{Ellis}},
  \bibinfo{author}{\bibfnamefont{G.}~\bibnamefont{Ridolfi}}, \bibnamefont{and}
  \bibinfo{author}{\bibfnamefont{F.}~\bibnamefont{Zwirner}},
  \bibinfo{journal}{Phys. Lett.} \textbf{\bibinfo{volume}{B257}},
  \bibinfo{pages}{83} (\bibinfo{year}{1991}).

\bibitem[{\citenamefont{Haber and Hempfling}(1993)}]{Haber:1993an}
\bibinfo{author}{\bibfnamefont{H.~E.} \bibnamefont{Haber}} \bibnamefont{and}
  \bibinfo{author}{\bibfnamefont{R.}~\bibnamefont{Hempfling}},
  \bibinfo{journal}{Phys. Rev.} \textbf{\bibinfo{volume}{D48}},
  \bibinfo{pages}{4280} (\bibinfo{year}{1993}), \eprint{hep-ph/9307201}.

\bibitem[{\citenamefont{Haber and Hempfling}(1991)}]{Haber:1990aw}
\bibinfo{author}{\bibfnamefont{H.~E.} \bibnamefont{Haber}} \bibnamefont{and}
  \bibinfo{author}{\bibfnamefont{R.}~\bibnamefont{Hempfling}},
  \bibinfo{journal}{Phys. Rev. Lett.} \textbf{\bibinfo{volume}{66}},
  \bibinfo{pages}{1815} (\bibinfo{year}{1991}).

\bibitem[{\citenamefont{Heinemeyer et~al.}(1998)\citenamefont{Heinemeyer,
  Hollik, and Weiglein}}]{Heinemeyer:1998kz}
\bibinfo{author}{\bibfnamefont{S.}~\bibnamefont{Heinemeyer}},
  \bibinfo{author}{\bibfnamefont{W.}~\bibnamefont{Hollik}}, \bibnamefont{and}
  \bibinfo{author}{\bibfnamefont{G.}~\bibnamefont{Weiglein}},
  \bibinfo{journal}{Phys. Lett.} \textbf{\bibinfo{volume}{B440}},
  \bibinfo{pages}{296} (\bibinfo{year}{1998}), \eprint{hep-ph/9807423}.

\bibitem[{\citenamefont{Chanowitz et~al.}(1978)\citenamefont{Chanowitz, Furman,
  and Hinchliffe}}]{Chanowitz:1978uj}
\bibinfo{author}{\bibfnamefont{M.~S.} \bibnamefont{Chanowitz}},
  \bibinfo{author}{\bibfnamefont{M.~A.} \bibnamefont{Furman}},
  \bibnamefont{and}
  \bibinfo{author}{\bibfnamefont{I.}~\bibnamefont{Hinchliffe}},
  \bibinfo{journal}{Phys. Lett.} \textbf{\bibinfo{volume}{B78}},
  \bibinfo{pages}{285} (\bibinfo{year}{1978}).

\bibitem[{\citenamefont{Lee et~al.}(1977)\citenamefont{Lee, Quigg, and
  Thacker}}]{Lee:1977eg}
\bibinfo{author}{\bibfnamefont{B.~W.} \bibnamefont{Lee}},
  \bibinfo{author}{\bibfnamefont{C.}~\bibnamefont{Quigg}}, \bibnamefont{and}
  \bibinfo{author}{\bibfnamefont{H.~B.} \bibnamefont{Thacker}},
  \bibinfo{journal}{Phys. Rev.} \textbf{\bibinfo{volume}{D16}},
  \bibinfo{pages}{1519} (\bibinfo{year}{1977}).

\bibitem[{\citenamefont{Peskin and Takeuchi}(1992)}]{Peskin:1991sw}
\bibinfo{author}{\bibfnamefont{M.~E.} \bibnamefont{Peskin}} \bibnamefont{and}
  \bibinfo{author}{\bibfnamefont{T.}~\bibnamefont{Takeuchi}},
  \bibinfo{journal}{Phys. Rev.} \textbf{\bibinfo{volume}{D46}},
  \bibinfo{pages}{381} (\bibinfo{year}{1992}).

\bibitem[{\citenamefont{Altarelli and Barbieri}(1991)}]{Altarelli:1990zd}
\bibinfo{author}{\bibfnamefont{G.}~\bibnamefont{Altarelli}} \bibnamefont{and}
  \bibinfo{author}{\bibfnamefont{R.}~\bibnamefont{Barbieri}},
  \bibinfo{journal}{Phys. Lett.} \textbf{\bibinfo{volume}{B253}},
  \bibinfo{pages}{161} (\bibinfo{year}{1991}).

\bibitem[{\citenamefont{Cho and Hagiwara}(2000)}]{Cho:1999km}
\bibinfo{author}{\bibfnamefont{G.-C.} \bibnamefont{Cho}} \bibnamefont{and}
  \bibinfo{author}{\bibfnamefont{K.}~\bibnamefont{Hagiwara}},
  \bibinfo{journal}{Nucl. Phys.} \textbf{\bibinfo{volume}{B574}},
  \bibinfo{pages}{623} (\bibinfo{year}{2000}), \eprint{hep-ph/9912260}.

\bibitem[{\citenamefont{Dabelstein}(1995)}]{Dabelstein:1994hb}
\bibinfo{author}{\bibfnamefont{A.}~\bibnamefont{Dabelstein}},
  \bibinfo{journal}{Z. Phys.} \textbf{\bibinfo{volume}{C67}},
  \bibinfo{pages}{495} (\bibinfo{year}{1995}), \eprint{hep-ph/9409375}.

\bibitem[{\citenamefont{Drees et~al.}(1992)\citenamefont{Drees, Hagiwara, and
  Yamada}}]{Drees:1991zk}
\bibinfo{author}{\bibfnamefont{M.}~\bibnamefont{Drees}},
  \bibinfo{author}{\bibfnamefont{K.}~\bibnamefont{Hagiwara}}, \bibnamefont{and}
  \bibinfo{author}{\bibfnamefont{A.}~\bibnamefont{Yamada}},
  \bibinfo{journal}{Phys. Rev.} \textbf{\bibinfo{volume}{D45}},
  \bibinfo{pages}{1725} (\bibinfo{year}{1992}).

\bibitem[{\citenamefont{Drees and Hagiwara}(1990)}]{Drees:1990dx}
\bibinfo{author}{\bibfnamefont{M.}~\bibnamefont{Drees}} \bibnamefont{and}
  \bibinfo{author}{\bibfnamefont{K.}~\bibnamefont{Hagiwara}},
  \bibinfo{journal}{Phys. Rev.} \textbf{\bibinfo{volume}{D42}},
  \bibinfo{pages}{1709} (\bibinfo{year}{1990}).

\bibitem[{\citenamefont{He et~al.}(2001)\citenamefont{He, Polonsky, and
  Su}}]{He:2001tp}
\bibinfo{author}{\bibfnamefont{H.-J.} \bibnamefont{He}},
  \bibinfo{author}{\bibfnamefont{N.}~\bibnamefont{Polonsky}}, \bibnamefont{and}
  \bibinfo{author}{\bibfnamefont{S.-f.} \bibnamefont{Su}},
  \bibinfo{journal}{Phys. Rev.} \textbf{\bibinfo{volume}{D64}},
  \bibinfo{pages}{053004} (\bibinfo{year}{2001}), \eprint{hep-ph/0102144}.

\bibitem[{\citenamefont{Haber}(1993)}]{Haber:1993wf}
\bibinfo{author}{\bibfnamefont{H.~E.} \bibnamefont{Haber}}
  (\bibinfo{year}{1993}), \eprint{hep-ph/9306207}.

\bibitem[{\citenamefont{Djouadi}(2008)}]{Djouadi:2005gj}
\bibinfo{author}{\bibfnamefont{A.}~\bibnamefont{Djouadi}},
  \bibinfo{journal}{Phys. Rept.} \textbf{\bibinfo{volume}{459}},
  \bibinfo{pages}{1} (\bibinfo{year}{2008}), \eprint{hep-ph/0503173}.

\bibitem[{\citenamefont{Heinemeyer et~al.}(2006)\citenamefont{Heinemeyer,
  Hollik, and Weiglein}}]{Heinemeyer:2004gx}
\bibinfo{author}{\bibfnamefont{S.}~\bibnamefont{Heinemeyer}},
  \bibinfo{author}{\bibfnamefont{W.}~\bibnamefont{Hollik}}, \bibnamefont{and}
  \bibinfo{author}{\bibfnamefont{G.}~\bibnamefont{Weiglein}},
  \bibinfo{journal}{Phys. Rept.} \textbf{\bibinfo{volume}{425}},
  \bibinfo{pages}{265} (\bibinfo{year}{2006}), \eprint{hep-ph/0412214}.

\bibitem[{\citenamefont{Hagiwara et~al.}(1994)\citenamefont{Hagiwara,
  Matsumoto, Haidt, and Kim}}]{Hagiwara:1994pw}
\bibinfo{author}{\bibfnamefont{K.}~\bibnamefont{Hagiwara}},
  \bibinfo{author}{\bibfnamefont{S.}~\bibnamefont{Matsumoto}},
  \bibinfo{author}{\bibfnamefont{D.}~\bibnamefont{Haidt}}, \bibnamefont{and}
  \bibinfo{author}{\bibfnamefont{C.~S.} \bibnamefont{Kim}},
  \bibinfo{journal}{Z. Phys.} \textbf{\bibinfo{volume}{C64}},
  \bibinfo{pages}{559} (\bibinfo{year}{1994}), \eprint{hep-ph/9409380}.

\bibitem[{\citenamefont{Erler and Langacker}(2010)}]{Erler:2010sk}
\bibinfo{author}{\bibfnamefont{J.}~\bibnamefont{Erler}} \bibnamefont{and}
  \bibinfo{author}{\bibfnamefont{P.}~\bibnamefont{Langacker}},
  \bibinfo{journal}{Phys. Rev. Lett.} \textbf{\bibinfo{volume}{105}},
  \bibinfo{pages}{031801} (\bibinfo{year}{2010}), \eprint{1003.3211}.

\bibitem[{\citenamefont{Kniehl}(1991)}]{Kniehl:1990mq}
\bibinfo{author}{\bibfnamefont{B.~A.} \bibnamefont{Kniehl}},
  \bibinfo{journal}{Nucl. Phys.} \textbf{\bibinfo{volume}{B352}},
  \bibinfo{pages}{1} (\bibinfo{year}{1991}).

\bibitem[{\citenamefont{Haber and Logan}(2000)}]{Haber:1999zh}
\bibinfo{author}{\bibfnamefont{H.~E.} \bibnamefont{Haber}} \bibnamefont{and}
  \bibinfo{author}{\bibfnamefont{H.~E.} \bibnamefont{Logan}},
  \bibinfo{journal}{Phys. Rev.} \textbf{\bibinfo{volume}{D62}},
  \bibinfo{pages}{015011} (\bibinfo{year}{2000}), \eprint{hep-ph/9909335}.

\end{thebibliography}

\end{document}